\input{aipcheck}

\documentclass[
    ,final            % use final for the camera ready runs
  ]
  {aipproc}

\layoutstyle{6x9}

\begin{document}

\title{Weak Interaction Studies with $^6$He}


\def\UW{Department of Physics and Center for Experimental Nuclear Physics and Astrophysics, University of Washington, Seattle, Washington 98195, USA}
\def\ANL{Physics Division, Argonne National Laboratory, Argonne, Illinois 60439, USA}
\def\TAM{Cyclotron Institute, Texas A\&M University, College Station, Texas 77843, USA}
\def\MSU{Department of Physics and Astronomy and National Superconducting Cyclotron Laboratory, Michigan State University, East Lansing, MI 48824, USA}
\def\LPC{LPC-Caen, ENSICAEN, Universit\'e de Caen, CNRS/IN2P3-ENSI, Caen, France}

\author{A.~Knecht}{
  address={\UW}
}
\author{Z.\,T.~Alexander}{
  address={\UW}
}
\author{Y.~Bagdasarova}{
  address={\UW}
}
\author{T.\,M.~Cope}{
  address={\UW}
}
\author{B.\,G.~Delbridge}{
  address={\UW}
}
\author{X.~Fl\'echard}{
  address={\LPC}
}
\author{A.~Garc\'ia}{
  address={\UW}
}
\author{R.~Hong}{
  address={\UW}
}
\author{E.~Li\'enard}{
  address={\LPC}
}
\author{P.~Mueller}{
  address={\ANL}
}
\author{O.~Naviliat-Cuncic}{
  address={\MSU}
}
\author{A.\,S.\,C.~Palmer}{
  address={\UW}
}
\author{R.\,G.\,H.~Robertson}{
  address={\UW}
}
\author{D.\,W.~Storm}{
  address={\UW}
}
\author{H.\,E.~Swanson}{
  address={\UW}
}
\author{S.~Utsuno}{
  address={\UW}
}
\author{F.~Wauters}{
  address={\UW}
}
\author{W.~Williams}{
  address={\ANL}
}
\author{C.~Wrede}{
  address={\MSU}
}
\author{D.\,W.~Zumwalt}{
  address={\UW}
}

\begin{abstract}
The $^6$He nucleus is an ideal candidate to study the weak interaction. To this end we have built a high-intensity source of $^6$He delivering $\sim$$10^{10}$ atoms/s to experiments. Taking full advantage of that available intensity we have performed a high-precision measurement of the $^6$He half-life that directly probes the axial part of the nuclear Hamiltonian. Currently, we are preparing a measurement of the beta-neutrino angular correlation in $^6$He beta decay that will allow to search for new physics beyond the Standard Model in the form of tensor currents. 
\end{abstract}

\maketitle

%%%%%%%%%%%%%%%%%%%%%%%%%%%%%%%%%%%%%%%%%%%%
%% MAINMATTER
%%%%%%%%%%%%%%%%%%%%%%%%%%%%%%%%%%%%%%%%%%%%

\section{Introduction}
Electroweak currents are at the basis of many interesting, fundamental processes such as, e.g., solar fusion, neutrino interactions, or muon and pion interactions \cite{Kub10}. As not all those processes are easily accessible directly, studies of nuclear beta decays can be used instead. Especially, measurements of half-lives and correlation coefficients have a long standing history in pinning down the underlying mechanism of the weak interaction.

The $^6$He nucleus in particular is a very attractive system for such studies:
\begin{itemize}
\item The $^6$He nucleus decays through a pure Gamow-Teller transition to the ground-state of $^6$Li with a 100\% branching ratio (except for a small beta-delayed deuteron emission branch of $\sim$$10^{-6}$ \cite{Raa09}).
\item The half-life of $\sim$807~ms allows enough time to manipulate the atoms while ensuring high decay rates.
\item The $^6$He nucleus is a simple few-body system for which \textit{ab initio}  calculations can be performed with precision (see, e.g., Refs.~\cite{Sch02,Nav03,Per07,Vai09}).
\item The $^6$He mass has recently been measured in a Penning trap \cite{Bro12} providing a reliable and high-precision $Q$-value of 3505.208(53)~keV.
\item With helium being a noble gas, effects due to chemical interactions are minimized.
\end{itemize}

In this paper, we will briefly present our $^6$He production source,  our measurement of the $^6$He half-life and the current status of the measurement of the beta-neutrino angular correlation in $^6$He beta decays.
 
\section{Production of $^6$He}
A detailed description of our high-intensity source for $^6$He can be found in Ref.~\cite{Kne11}. We produce $^6$He using the tandem Van de Graaff accelerator available at the Center for Experimental Nuclear Physics and Astrophysics of the University of Washington. A deuteron beam impinges on molten lithium held in a stainless steel cup producing $^6$He primarily via the reaction $^7$Li($^2$H, $^3$He)$^6$He. The $^6$He atoms subsequently diffuse out into vacuum and are transported to the experimental area by means of a turbo-molecular pump.

While we reported in Ref.~\cite{Kne11} extracted $^6$He rates in the experimental area of $10^9$~atoms/s, we have in the meantime reached values in excess of $10^{10}$~atoms/s by increasing the deuteron beam currents up to 14~$\mu$A and improving our beam and production diagnostics. This is currently the highest intensity of $^6$He atoms available for experiments world-wide and even surpasses the $^6$He intensities reached in the 1960s at high power nuclear reactors \cite{Ple64}.

\section{Measurement of the $^6$He half-life}
The pure Gamow-Teller decay of the $^6$He decay allows to probe axial currents in a simple nuclear system. Contrary to the vector currents that are protected by the Conserved Vector Current hypothesis, axial currents could be changed in the nuclear medium leading to a renormalization of the axial coupling constant found in the decay of free neutrons. Several works have calculated the $^6$He half-life using \textit{ab initio} methods \cite{Sch02,Nav03,Per07,Vai09} and found agreement between the experimental and theoretical matrix elements at the percent level allowing for only minimal ``quenching'' of the axial coupling constant \cite{Cho93, Cau95}. 

With theoretical calculations pushing towards precisions of one percent and below and with previous half-life measurements being discrepant we decided to perform a high-precision measurement of the $^6$He half-life that was recently published in Ref.~\cite{Kne12}. The $^6$He atoms from our source were transported to the experimental area and confined in a stainless steel cylinder closed off on one side by a valve and on the other by a thin copper window. In front of the copper window we had mounted two scintillators detecting the decay electrons in coincidence. We took data in cycles in which we filled the counting volume with $^6$He for 8~s, closed it off and then counted the decay electrons for 16~s. In order to study systematic effects we performed the measurements at various initial decay rates, checked for helium diffusion effects by inserting a stainless steel rod into the counting volume, and performed background runs with the inlet valve closed. In addition, several offline measurements were performed to determine dead times, after pulsing, and other effects .

\begin{figure}
  \centering
  \begin{tabular}{lr}
  \includegraphics[width=0.45\textwidth]{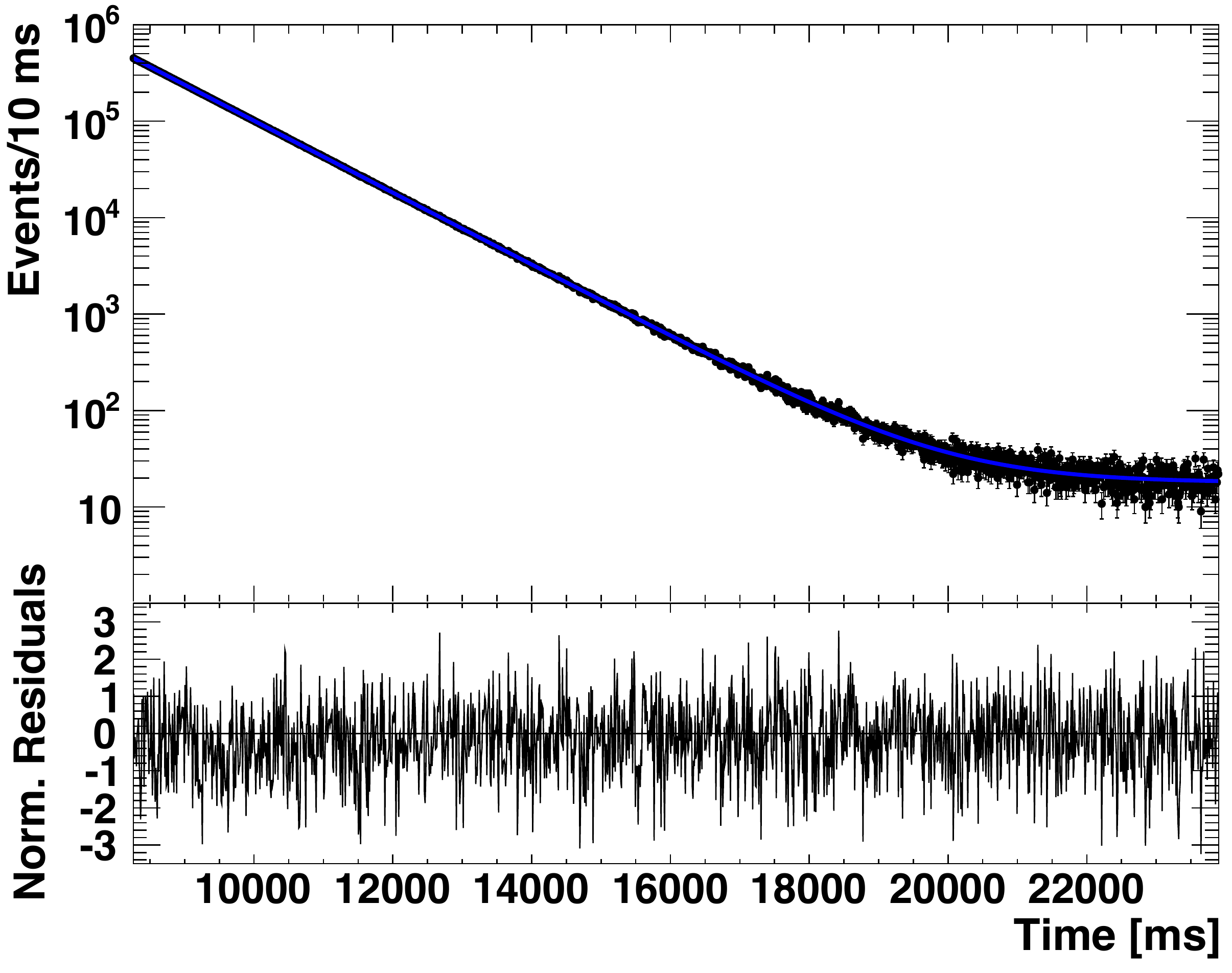}
&
 \includegraphics[width=0.45\textwidth]{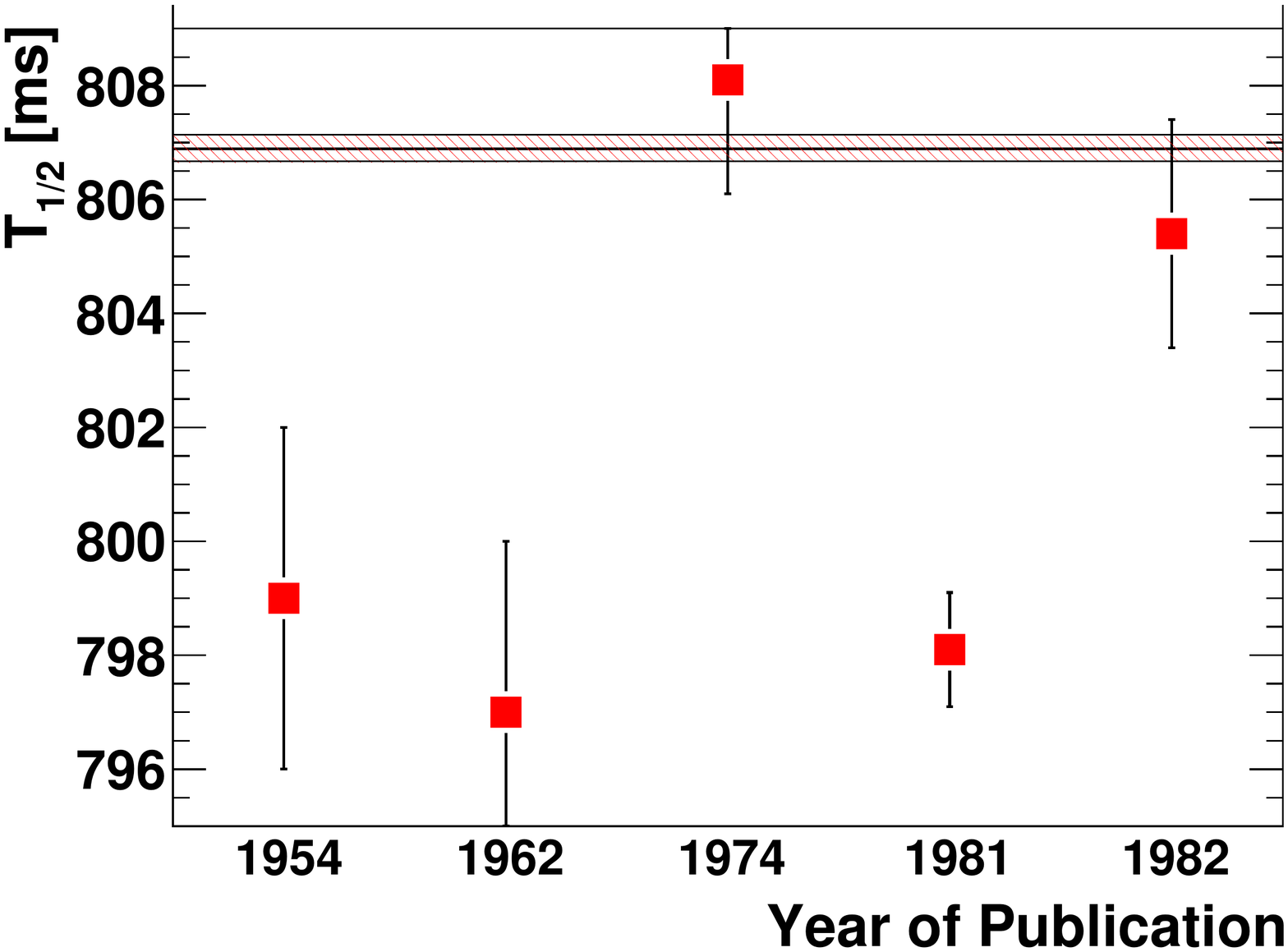} \\
 \end{tabular}
  \caption{Left: $^6$He decay curve data including a fit and the corresponding residuals. Right: Compilation of the previous five precision measurements of the $^6$He half-life \cite{Kli54,Bie62,Wil74,Bar81,Alb82} with the dashed red band depicting the value for the $^6$He half-life obtained in our measurement. Both figures are taken from Ref.~\cite{Kne12}.}
  \label{6He_half-life_plots}
\end{figure}

Figure~\ref{6He_half-life_plots} (left) shows the summed decay curve for part of the data together with a fit and its corresponding residuals. Due to the coincidence requirement between the two scintillators the background rate is very low allowing to follow the decay curve over four orders of magnitude. We performed a careful analysis of possible systematic effects presented in Refs.~\cite{Kne12, Kne12b}. The final result for the $^6$He half-life is:
\begin{equation}
\mathrm{T}_{1/2} = 806.89 \pm 0.11_\mathrm{stat}  \,^{+0.23}_{-0.19} \,_\mathrm{syst}\;\mathrm{ms}
\end{equation}

The previous five precision results of the $^6$He half-life \cite{Kli54,Bie62,Wil74,Bar81,Alb82} are shown together with our result in Fig.~\ref{6He_half-life_plots} (right). There is a clear discrepancy between our result and three of the previous results. Because the possibility of diffusion out of the target was not explicitly addressed in these experiments, we suspect that this may be the cause of the discrepancy. 

From our measurement we calculated the $ft$-value to be $803.04^{+0.26}_{-0.23}$~s and extracted the experimental Gamow-Teller matrix element $| M_\mathrm{GT} | = 2.1645(43)$ using the value for the axial coupling constant obtained from free neutron decay $g_A = -1.2701(25)$~\cite{Nak10} providing a precise and reliable value to be used in comparison with theoretical calculations. 

\section{Towards a measurement of the $^6$He beta-neutrino angular correlation}
A measurement of the beta-neutrino angular correlation probes the symmetry of the underlying interaction. In the case of $^6$He -- a pure Gamow-Teller decay -- the interaction is purely axial in nature according to the Standard Model. Testing for deviations from the Standard Model prediction can uncover contributions from new physics in the form of tensor currents. A review of the status of the field can be found in Refs.~\cite{Sev06,Sev11}. The current best measurement for $^6$He -- reaching a precision of 0.9\% -- was performed in 1963 by measuring the energy distribution of the recoil ions \cite{Joh63}. A 3.1\%-precision value for the correlation coefficient from a coincidence measurement of the emitted beta particle and recoil ion using LPCTrap at GANIL has been reported recently \cite{Fle11} with data to reach a statistical accuracy of 0.7\% currently under analysis.

\begin{figure}
  \centering
  \begin{tabular}{lr}
  \includegraphics[width=0.7\textwidth]{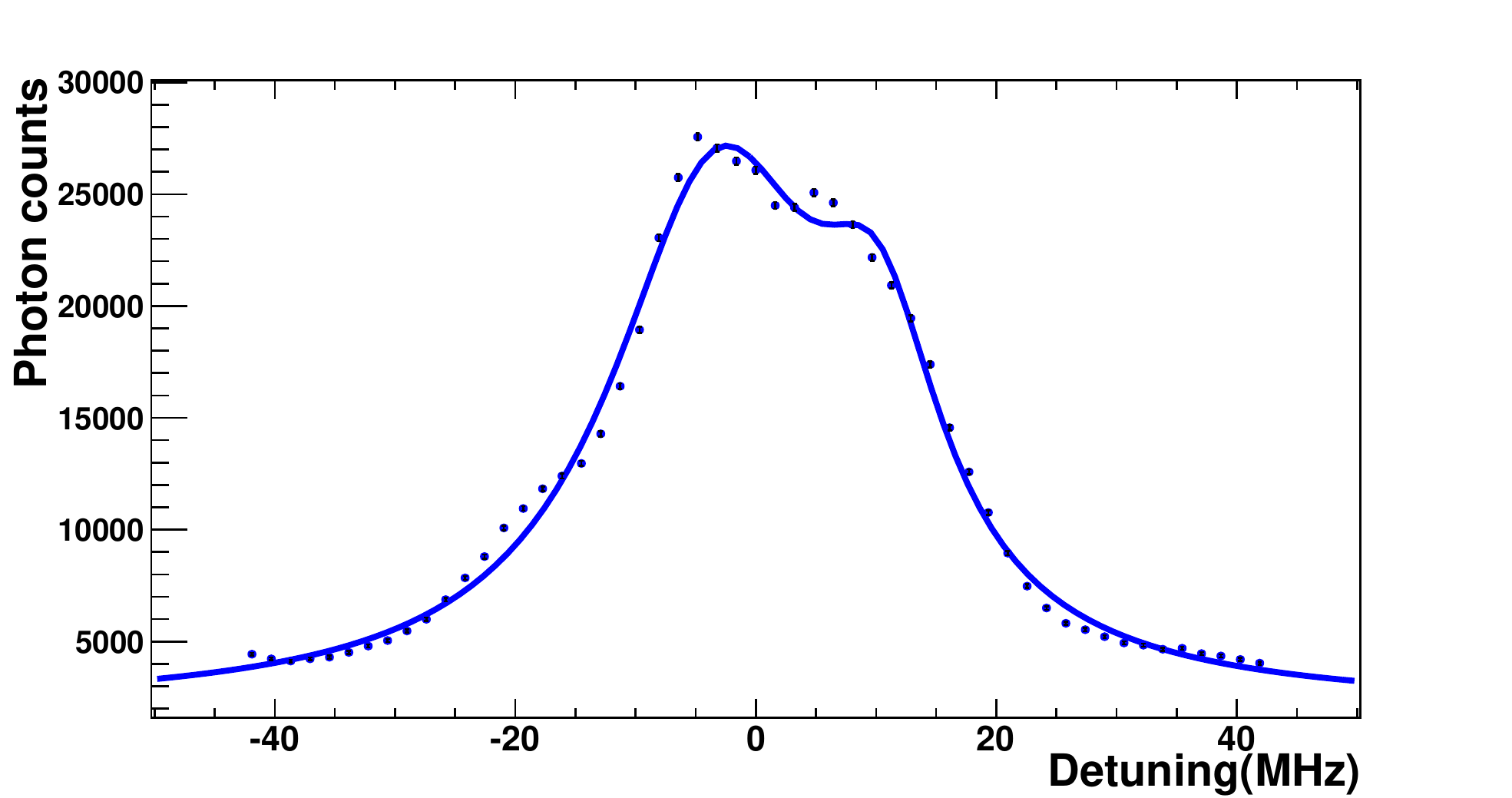}
&
 \includegraphics[width=0.28\textwidth]{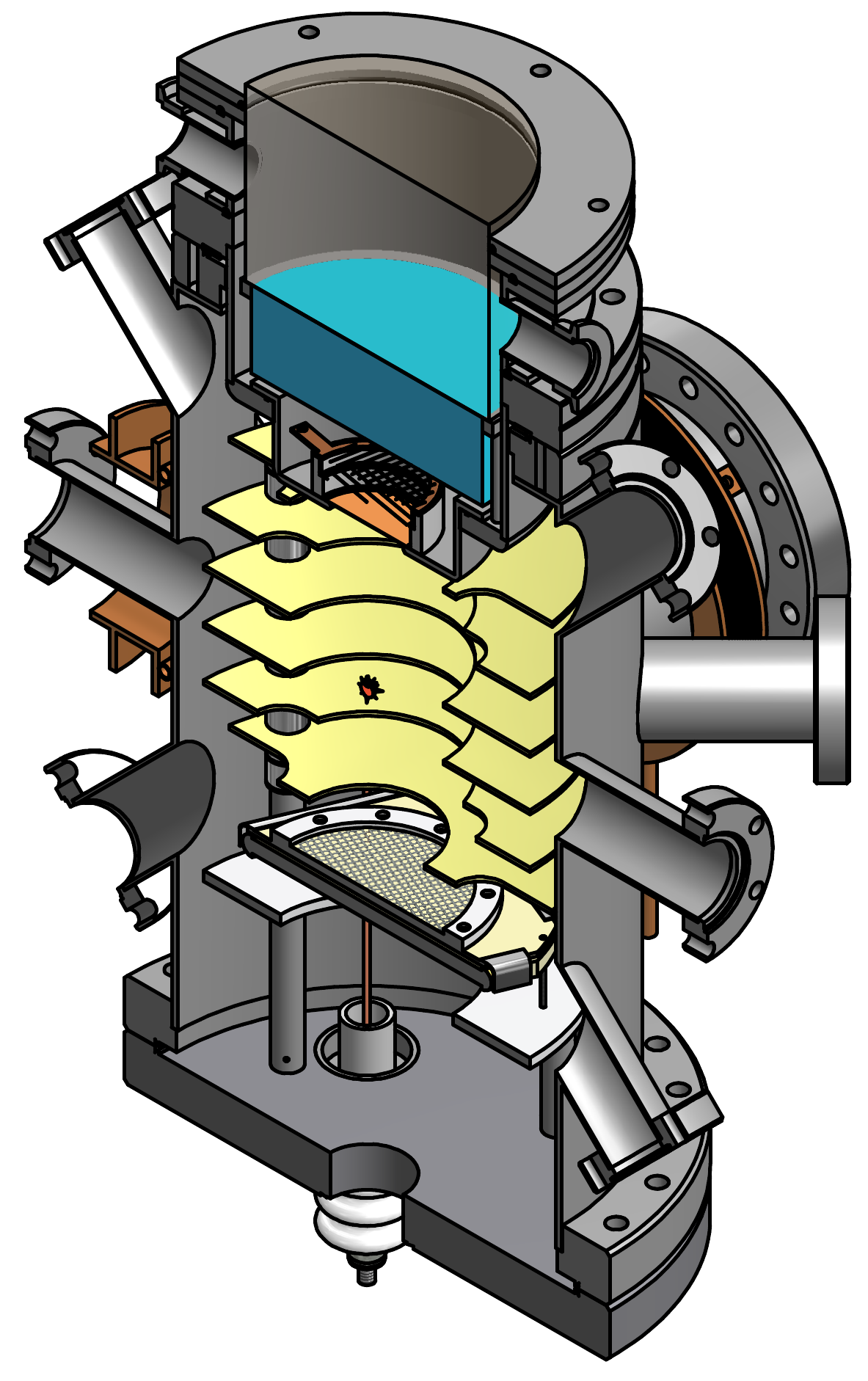} \\
 \end{tabular}
  \caption{Left: Fluorescence photon counts as a function of the wavelength of a laser probing the trapped $^6$He atoms. Right: Cross-sectional view of the detection MOT with a MCP on the bottom, a 2~kV/cm electrode structure in the middle, and a proportional counter and thick scintillator on the top. The $^6$He atom cloud is located in the center of the electrode structure.}
  \label{6He_traps}
\end{figure}

The $^6$He atoms coming from our source are extracted into the experimental area and pass through an RF discharge source to get excited into the metastable $2s$-state. At this point we can access the $2s$-$2p$ cycling transition using a laser at 1083~nm. The metastable $^6$He atoms are subsequently transversely cooled, focused, slowed down and finally trapped in a magneto-optical trap (MOT). We detect the number of atoms in the trap by tuning a second laser onto the $2p$-$3s$ resonance at 706~nm and collecting the fluorescence signal with a photomultiplier tube (PMT). Figure~\ref{6He_traps} (left) shows the number of photons detected by the PMT as the wavelength of the probe laser is varied. We have achieved a total number of 500 atoms in our trap and still have ways to increase the number of atoms to our design goal of 1000 trapped atoms.

Currently we are working on instrumenting a second MOT with the necessary detectors in order to register the emitted beta particle and recoil ion in coincidence. The $^6$He atoms will be transferred from the trapping MOT through a small opening to the detection MOT in order to reduce the amount of background present from untrapped atoms. On one side of the detector chamber we will mount a micro-channel plate (MCP) detector that records the position and time-of-flight of the recoil ions (with respect to the detection of the beta particle). An electrode structure providing an electric field of 2~kV/cm will collect and accelerate all the ions onto the MCP. On the other side beta particles passing through a thin beryllium window will be detected by a $\Delta$E-E detector configuration consisting of a proportional counter and a thick scintillator read out by a PMT. The setup can be seen in Fig.~\ref{6He_traps} (right). In this configuration we are aiming at performing a 1\% measurement within the next year. At a later stage we will change the detection MOT to a dipole trap allowing for better control of the atom cloud and better access for particle detectors. This should then lead to the final anticipated precision of 0.1\%.

\section{Summary}
The $^6$He nucleus is an ideal system to probe the weak interaction in detail. To this end we have developed a high-intensity source delivering $\sim$$10^{10}$ $^6$He atoms/s to experiments.

We have performed a high-precision measurement of its half-life that allows for improved comparisons between experiment and \textit{ab initio} theory calculation, deepening our understanding of the nuclear Hamiltonian.

With 500 $^6$He atoms trapped in a MOT we are ready to pursue the next steps in our project of measuring the beta-neutrino angular correlation. This will allow for searches for new physics beyond the Standard Model in the form of tensor currents.

%%%%%%%%%%%%%%%%%%%%%%%%%%%%%%%%%%%%%%%%%%%%%%%%
%% BACKMATTER
%%%%%%%%%%%%%%%%%%%%%%%%%%%%%%%%%%%%%%%%%%%%%%%%

\begin{theacknowledgments}
We thank the excellent staff at the Center for Experimental Nuclear Physics and Astrophysics. This work has been supported by the US Department of Energy under DE-FG02-97ER41020 and under Contract No. DEAC02-06CH11357.

\end{theacknowledgments}


\begin{thebibliography}{23}
\expandafter\ifx\csname natexlab\endcsname\relax\def\natexlab#1{#1}\fi
\providecommand{\enquote}[1]{``#1''}
\expandafter\ifx\csname url\endcsname\relax
  \def\url#1{\texttt{#1}}\fi
\expandafter\ifx\csname urlprefix\endcsname\relax\def\urlprefix{URL }\fi
\providecommand{\eprint}[2][]{\url{#2}}

\bibitem[Kubodera and Rho(2011)]{Kub10}
K.~Kubodera, and M.~Rho, \enquote{{Effective Field Theory and High-Precision
  Calculations of Nuclear Electroweak Processes},} in \emph{{From Nuclei to
  Stars, Festschrift in Honor of Gerald E. Brown}}, edited by S.~Lee, World
  Scientific, Singapore, 2011.

\bibitem[Raabe et~al.(2009)]{Raa09}
R.~Raabe, et~al., \emph{Phys. Rev. C} \textbf{80}, 054307 (2009).

\bibitem[Schiavilla and Wiringa(2002)]{Sch02}
R.~Schiavilla, and R.~B. Wiringa, \emph{Phys. Rev. C} \textbf{65}, 054302
  (2002).

\bibitem[Navr\'atil and Ormand(2003)]{Nav03}
P.~Navr\'atil, and W.~E. Ormand, \emph{Phys. Rev. C} \textbf{68}, 034305
  (2003).

\bibitem[Pervin et~al.(2007)]{Per07}
M.~Pervin, S.~C. Pieper, and R.~B. Wiringa, \emph{Phys. Rev. C} \textbf{76},
  064319 (2007).

\bibitem[Vaintraub et~al.(2009)]{Vai09}
S.~Vaintraub, N.~Barnea, and D.~Gazit, \emph{Phys. Rev. C} \textbf{79}, 065501
  (2009).

\bibitem[Brodeur et~al.(2012)]{Bro12}
M.~Brodeur, et~al., \emph{Phys. Rev. Lett.} \textbf{108}, 052504 (2012).

\bibitem[Knecht et~al.(2011)]{Kne11}
A.~Knecht, et~al., \emph{Nucl. Instr. Method in Phys. Res. A} \textbf{660}, 43
  (2011).

\bibitem[Pleasonton and Johnson(1964)]{Ple64}
F.~Pleasonton, and C.~H. Johnson, \emph{Rev. Sci. Instrum.} \textbf{35}, 97
  (1964).

\bibitem[Chou et~al.(1993)]{Cho93}
W.-T. Chou, E.~K. Warburton, and B.~A. Brown, \emph{Phys. Rev. C} \textbf{47},
  163 (1993).

\bibitem[Caurier et~al.(1995)]{Cau95}
E.~Caurier, A.~Poves, and A.~P. Zuker, \emph{Phys. Rev. Lett.} \textbf{74},
  1517 (1995).

\bibitem[Knecht et~al.(2012{\natexlab{a}})]{Kne12}
A.~Knecht, et~al., \emph{Phys. Rev. Lett.} \textbf{108}, 122502
  (2012{\natexlab{a}}).

\bibitem[Knecht et~al.(2012{\natexlab{b}})]{Kne12b}
A.~Knecht, et~al., \emph{Phys. Rev. C}  (2012{\natexlab{b}}), accepted for
  publication.

\bibitem[Kline and Zaffarano(1954)]{Kli54}
R.~M. Kline, and D.~J. Zaffarano, \emph{Phys. Rev.} \textbf{96}, 1620 (1954).

\bibitem[Bienlein and Pleasonton(1962)]{Bie62}
J.~K. Bienlein, and F.~Pleasonton, \emph{Nuclear Physics} \textbf{37}, 529
  (1962).

\bibitem[Wilkinson and Alburger(1974)]{Wil74}
D.~H. Wilkinson, and D.~E. Alburger, \emph{Phys. Rev. C} \textbf{10}, 1993
  (1974).

\bibitem[Barker et~al.(1981)]{Bar81}
P.~H. Barker, T.~B. Ko, and M.~J. Scandle, \emph{Nuclear Physics A}
  \textbf{372}, 45 (1981).

\bibitem[Alburger(1982)]{Alb82}
D.~E. Alburger, \emph{Phys. Rev. C} \textbf{26}, 252 (1982).

\bibitem[Nakamura et~al.(2010)]{Nak10}
K.~Nakamura, et~al., \emph{J. Phys. G} \textbf{37}, 075021 (2010).

\bibitem[Severijns et~al.(2006)]{Sev06}
N.~Severijns, M.~Beck, and O.~Naviliat-Cuncic, \emph{Rev. Mod. Phys.}
  \textbf{78}, 991 (2006).

\bibitem[Severijns and Naviliat-Cuncic(2011)]{Sev11}
N.~Severijns, and O.~Naviliat-Cuncic, \emph{Annual Review of Nuclear and
  Particle Science} \textbf{61}, 23 (2011).

\bibitem[Johnson et~al.(1963)]{Joh63}
C.~H. Johnson, F.~Pleasonton, and T.~A. Carlson, \emph{Phys. Rev.}
  \textbf{132}, 1149 (1963).

\bibitem[Fl\'echard et~al.(2011)]{Fle11}
X.~Fl\'echard, et~al., \emph{Journal of Physics G: Nuclear and Particle
  Physics} \textbf{38}, 055101 (2011).

\end{thebibliography}
\end{document}